\documentclass[aps,prl,twocolumn,floatfix]{revtex4}
\pdfoutput=1
\usepackage{amsmath}
\usepackage{graphicx}
\usepackage{latexsym}
\usepackage{amsfonts}
\usepackage{amssymb}

\usepackage{colordvi}

\newcommand{\bra}[1]{\ensuremath{\left< #1 \right|}}
\newcommand{\ket}[1]{\ensuremath{\left| #1 \right>}}

\begin{document}

\title{Adaptively measuring the temporal shape of ultrashort single photons for higher-dimensional
quantum information processing}

\author{C. Polycarpou$^{1,2}$, K. N. Cassemiro$^{3,4}$, G. Venturi$^{2}$, A.
Zavatta$^{1,2}$, and M. Bellini$^{1,2}$}

%\begin{document}
%\maketitle

\affiliation{$^{1}$Istituto Nazionale di Ottica (INO-CNR), L.go E. Fermi 6,
50125 Florence, Italy,\\
$^{2}$LENS and Department of Physics, University of Firenze, 50019 Sesto
Fiorentino, Florence, Italy,\\
$^{3}$Max Planck Institute for the Science of Light, G\"unther Scharowsky stra{\ss}e
1/Bau 24, 91058 Erlangen, Germany,\\
$^{4}$Departamento de F\'isica, Universidade Federal de Pernambuco, 50670-901 Recife -- PE, Brazil}

\begin{abstract}
A photon is the single excitation of a particular spatiotemporal mode of the electromagnetic field. A precise knowledge of the mode structure is therefore essential for its processing and detection, as well as for applying generic quantum light states to novel technologies. Here we demonstrate an adaptive scheme for reconstructing the arbitrary amplitude and phase spectro-temporal profile of an ultrashort single-photon pulse. The method combines techniques from the fields of ultrafast coherent control and quantum optics to map the mode of a fragile quantum state onto that of an intense coherent field. In addition, we show that the possibility of generating and detecting quantum states in multiple spectro-temporal modes may serve as a basis for encoding qubits (and qudits) into single, broadband, ultrashort, photons. Providing access to a much larger Hilbert space, this scheme may boost the capacity of current quantum information protocols.
\end{abstract}

\maketitle

%\newpage
Most of the experiments performed so far in the field of quantum optics have relied on the generation, manipulation, and detection, of single photons and other nonclassical light states in single, well-defined modes~\cite{bellini10}. Generation of quantum light in a superposition of two spatial or spectro-temporal (S/T) modes has been used to encode quantum information (e.g. time-bin schemes~\cite{brendel99,zavatta06}) or for metrological applications~\cite{boto00,dangelo01,banaszek09,giovannetti04}. However, simple single- or two-mode systems limit the capacity in the communication, manipulation and storage of quantum information and, equally importantly, are not able to seize the complexity of realistic states in the laboratory.

On one hand, the possibility of using multimode quantum states of light, i.e. a single beam holding several independent quantum channels, may offer several advantages. As much as wavelength multiplexing has revolutionized the area of optical telecommunications, multimode states might boost the complexity of quantum networks and enhance the execution of quantum information protocols~\cite{Bechmann-Pasquinucci00,cerf02,Collins02,durt04,vertesi10}. In the spatial domain multimode states have been used in quantum imaging applications~\cite{kobolov00,lugiato02,brida10,broadbent09,janousek09} and the orbital angular momentum of single photons has been explored~\cite{mair01,leach04}. In the spectral and temporal domain they have been proposed for cluster-state quantum computing~\cite{menicucci08,Pysher11} and for enhanced time metrology~\cite{fabre08}. The encoding of quantum information in the S/T degrees of freedom of a single photon has also been proposed for novel quantum cryptographic schemes, such as differential phase shift QKD~\cite{inoue02}.

On the other hand, the modes where the quantum states are prepared are often not perfectly coinciding with those used for their processing and detection, and this may significantly degrade the quality of all the possible applications based on such states. For example, future quantum networks require that light not only conveys information through optical links but also interacts with atomic species, allowing one to perform quantum processing and implement memory units~\cite{lvovsky09,reim10}. These tasks require a very specific and precise preparation of the photon wavepacket, i.e. its S/T mode, such that it optimally couples to the different possible interfaces. This goal has already motivated a series of theoretical proposals~\cite{vasilev10,kielpinski11} and experiments, where the capability of arbitrarily shaping the phase~\cite{specht09} and amplitude~\cite{kolchin08} profile of the mode of ``long" (typically hundreds of nanosecond) single photons has been shown.

Even the seemingly simple task of detecting a single photon may be completely hopeless if an accurate knowledge of the exact spatiotemporal mode occupied by the photon is not available in advance. In a typical experiment for the complete tomographic reconstruction of some quantum light state, homodyne detection only works with sufficient efficiency if the spatiotemporal mode of the reference classical coherent beam (the so-called local oscillator, LO) is perfectly matched to the signal mode~\cite{lvovsky09:rmp}. If little or no prior information on such signal mode is at hand, or if the mode itself has been somehow distorted during the propagation from the source to the detector, one may completely miss the target in the detection stage.

If we restrict ourselves to a single spatial mode and just examine the spectral and temporal degrees of freedom, a monochromatic single photon at frequency $\omega$ can be defined as
\begin{equation}
\ket{\omega} =\hat a^{\dag}(\omega)\ket{0}
\end{equation}
where $\hat a^{\dag}$ is the bosonic creation operator. A single photon in an arbitrary S/T mode can thus be written as
\begin{equation}
\ket{1}_{\Psi} = \int{d\omega \Psi(\omega)\ket{\omega}}
\end{equation}
and, unless the shape of the wave-packet defined by the complex amplitude $\Psi(\omega)$ is accurately known, any characterization or use of such a quantum state is doomed to fail or perform very poorly. Measuring and controlling $\Psi(\omega)$ is therefore essential.

Starting from the above simple and prototypical case, we have developed a novel scheme that combines techniques from the field of ultrafast coherent control and quantum optics to completely reconstruct $\Psi(\omega)$, namely the amplitude and phase profile of an ultrashort single-photon, by mapping it to that of an intense coherent field. More in detail, the shape of the LO in a pulsed homodyne detection system is first iteratively adjusted to best match the single-photon shape by means of an adaptive genetic algorithm based on the measured homodyne data; successively, the spectral amplitude and phase of the single photon are characterized by measurements on the optimized local oscillator. Besides demonstrating that this approach allows us to perform the full reconstruction of quantum states in unknown and arbitrarily-shaped modes, we also show that it may serve as a basis for novel quantum information protocols based on the encoding of qubits and qudits onto the temporal and the spectral degrees of freedom of single broadband ultrashort photons.

The experimental setup is schematically shown in Fig.1. We conditionally generate single photons by the process of spontaneous parametric down-conversion in a nonlinear crystal pumped by the second harmonic of a mode-locked 80-fs pulse train at 800 nm. Detection of a tightly filtered idler photon heralds the presence of a highly-pure single photon in the signal mode. An ultrafast, time-domain, homodyne detector~\cite{zavatta02:josab,zavatta06:lpl} is used to measure quadrature data using coherent LO pulses coming from the laser.
\begin{figure}[h]
\includegraphics [width=1\linewidth]{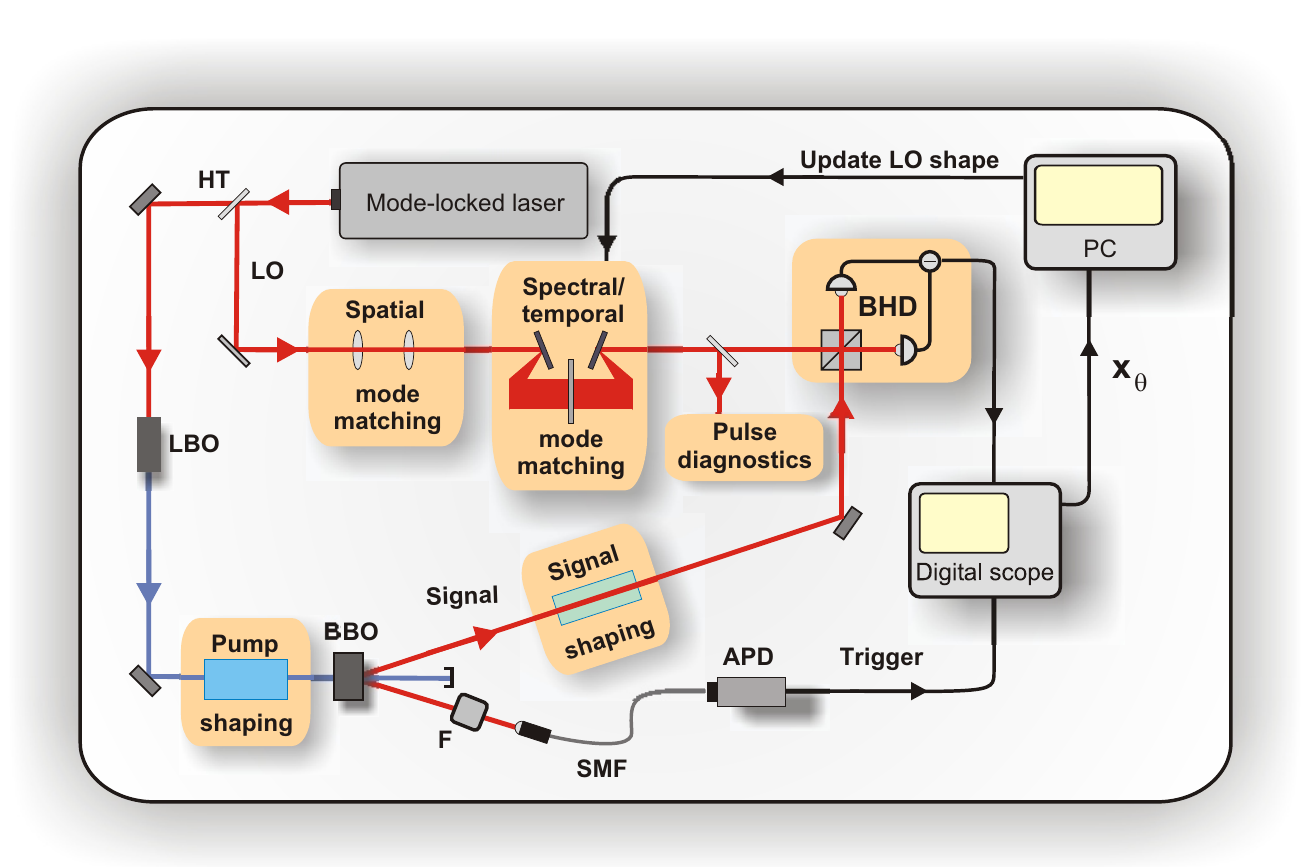}
\caption{\textbf{Experimental setup}. The output of the mode-locked femtosecond laser is frequency doubled in a lithium triborate (LBO) crystal and then used to pump spontaneous parametric down-conversion in a thin (300 $\mu m$) $\beta$-barium borate (BBO) crystal. The idler photons are detected by an avalanche silicon photodiode (APD, Perkin Elmer AQR-14) after narrow spectral filtering (F) and single-mode fiber (SMF) transmission. The signal field is conditionally measured by a balanced homodyne detector (BHD) triggered by the idler counts. The local oscillator (LO) is obtained by splitting a part of the laser emission with a high-transmission beam-splitter (HT), and is spatially mode-matched to the conditional single-photon mode by a combination of lenses. The S/T matching of the LO is performed by means of a spatial light modulator (CRI, SLM-128-D-VM) placed in the Fourier plane of a ``zero dispersion (4$f$) line''. The spectral and temporal (intensity and phase) profiles of the LO are measured with a spectrometer (Ocean Optics HR4000), with an interferometric autocorrelator, and with a second-harmonic-generation-based FROG device (FROG-Scan, MesaPhotonics). Pump and signal shaping are obtained by the use of a stabilized Michelson interferometer, and a 10-cm-long block of BK7 glass, respectively.}
\end{figure}

In general, the measured state is not pure, but rather a mixture of single photon and vacuum, that can be expressed as
\begin{equation}
\eta \ket{1}\bra{1} +(1-\eta)\ket{0} \bra{0}.\label{eq_eta}
\end{equation}
Apart from other systematic and relatively constant factors (like the limited efficiency in the detectors, optical losses, electronic noise, dark counts, and residual spatial mismatch), the S/T mode matching between the LO and the single photons has a direct impact on the measured efficiency $\eta$, which can thus be used as a merit parameter for optimizing the LO shape.

We use a spatial light modulator (SLM) placed in the Fourier plane of a ``zero dispersion (4$f$) line''~\cite{weiner11,monmayrant10} to shape the LO. Here, different pulse frequencies are mapped into different spatial positions, and the pixelated SLM is used to apply arbitrary amplitude and phase modulation to the different spectral components in response to applied voltages. The mapping of the single-photon S/T profile onto the LO pulse is done by means of an optimization procedure based on an evolutionary genetic algorithm~\cite{zeidler01,baumert97}.
\begin{figure}[h]
\includegraphics [width=1\linewidth]{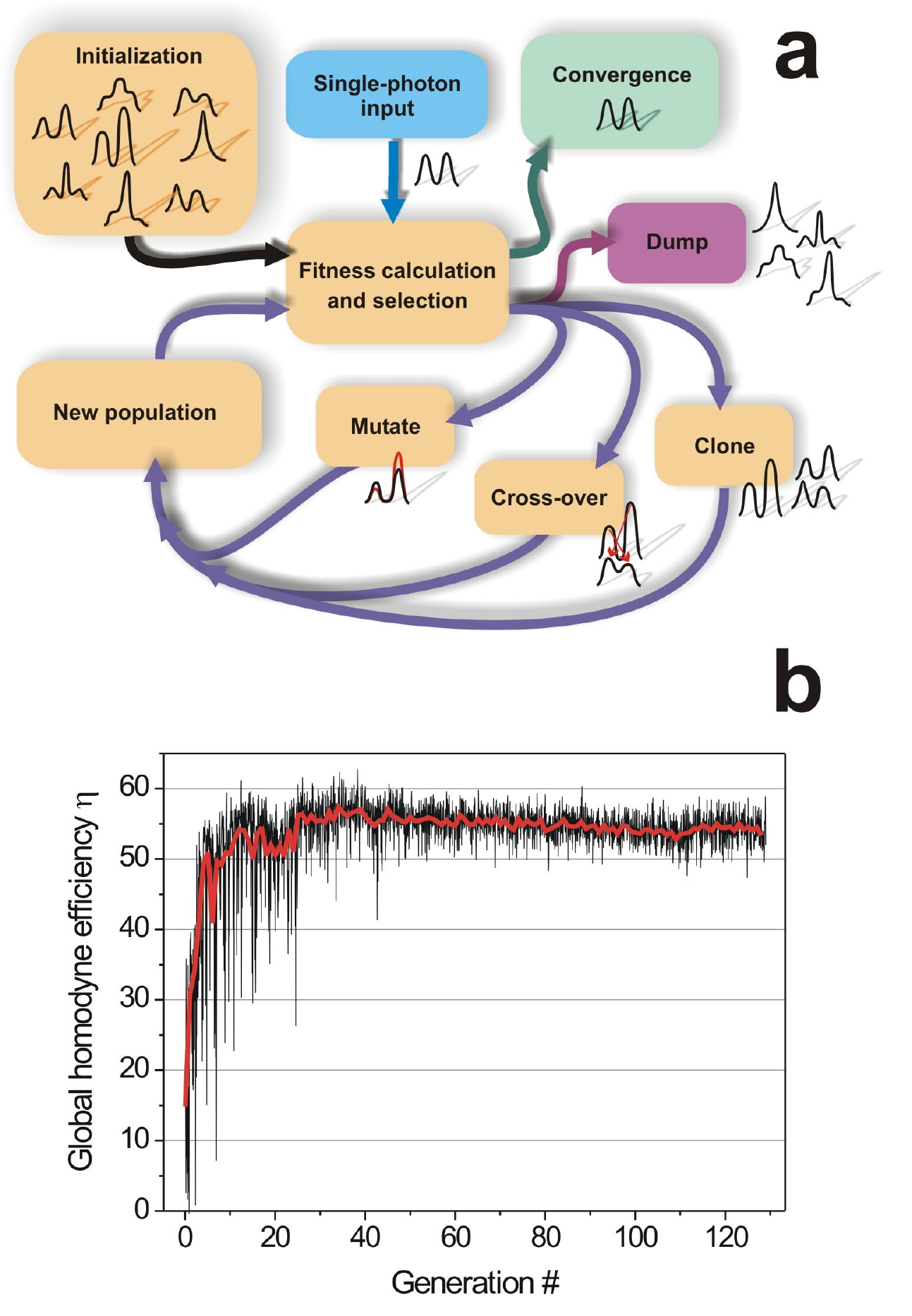}
\caption{\textbf{Searching for the photon shape}. a) Scheme of the genetic algorithm used to match the LO spectro-temporal shape to that of the generated single photons. b) Typical evolution of the fitness parameter (the total homodyne efficiency $\eta$) across several generations towards convergence to the optimal shape. The black curve corresponds to the different individuals, while the red one is the average efficiency in each generation. Convergence is obtained already after about 30-40 generations, then a slow decay is observed, probably due to deteriorating spatial mode-matching and decreasing laser power during the long (about 14 hours) measurement run. The sudden, isolated, drops in the efficiency curve are due to the appearance of ``bad" random mutations.}
\end{figure}
One first generates an initial population of random voltage profiles (the individuals). Each individual is sequentially applied to the SLM, thus shaping the LO pulse according to its particular configuration of voltage levels (the genes). Each shaped LO pulse is used for the homodyne detection of the single photon and the observed value of efficiency $\eta_i$ is assigned as the fitness parameter of that particular $i^{th}$ individual. The voltage sequences of the best individuals are then cloned and used to create the individuals of the next generation by means of cross-overs and mutations (see Fig.2a). The algorithm keeps producing new generations in an iterative way until convergence towards a steady optimum value $\eta_{opt}$ is reached, corresponding to the best possible mode-matching between the LO and the photon shape. Finally, we use a spectrometer, an interferometric autocorrelator, and a commercial FROG~\cite{monmayrant10} device to fully analyze the S/T intensity and phase profiles of the optimal LO pulses, a faithful representation of the single-photon ones.

The method is first applied to the heralded single photons directly coming from the parametric crystal. The optimal LO shape obtained for such un-modulated single photons is shown in Fig.3a. An almost flat spectral phase indicates the absence of any frequency chirp. In order to put the method to a more stringent test, we need to make use of arbitrarily-shaped single-photon pulses. Differently from other recent schemes~\cite{specht09, kolchin08} using ``long" single photons produced by narrow-band atomic samples, where electronic modulators and detectors can keep up with the slow temporal evolution of the photon wavepacket, here both the modulation and the detection stages have to use completely different approaches due to the ultrashort duration of the wavepackets.
\begin{figure}[h]
\includegraphics [width=.8\linewidth]{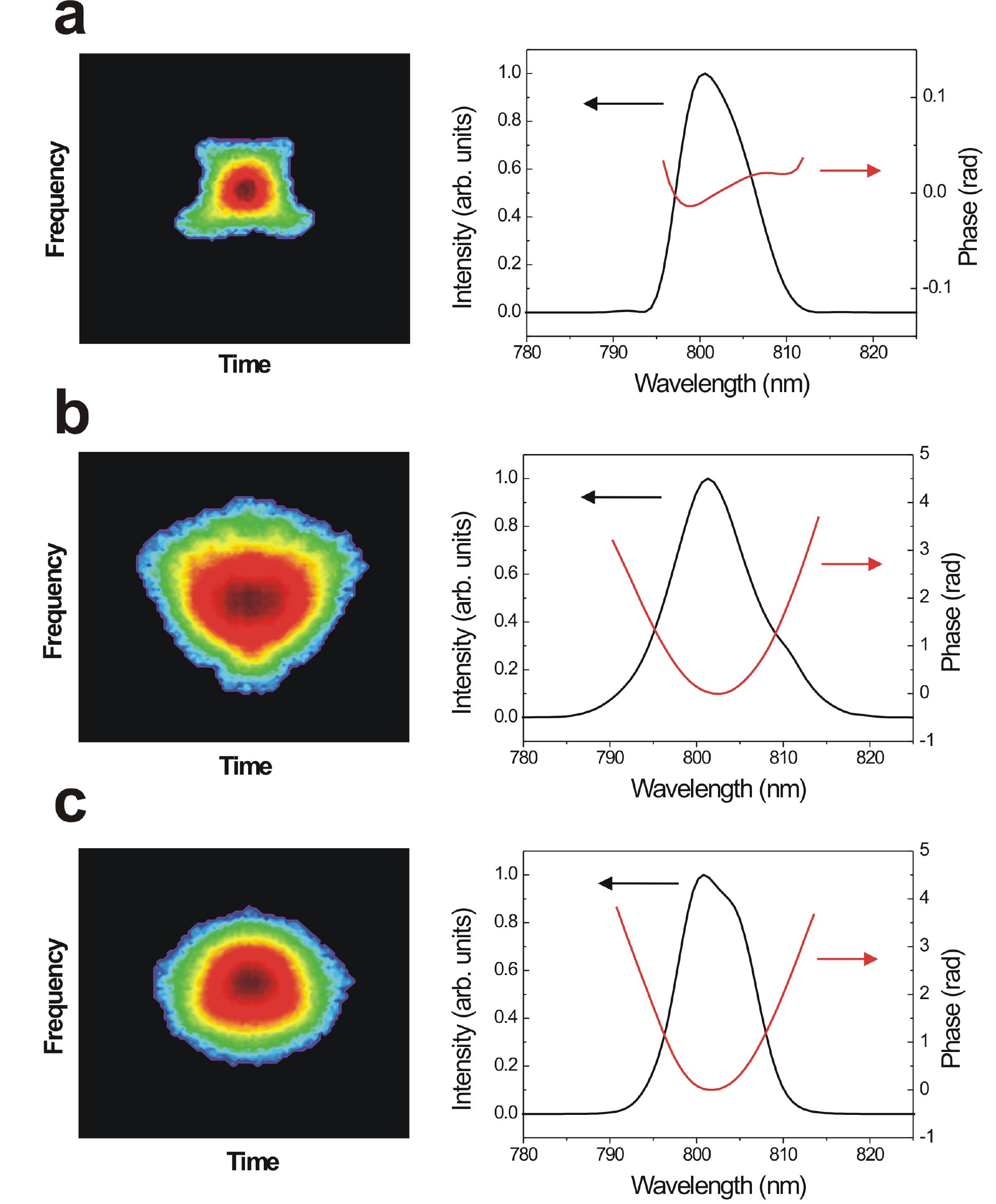}
\caption{\textbf{Frequency-dispersed single photon}. Experimental FROG traces (left panels) and reconstructed spectral intensity and phase profiles (right panels) for: a) optimal LO pulse matching the un-dispersed single photons, after a full amplitude and phase optimization; b) optimal LO pulse matching a frequency-dispersed single-photon after a phase-only optimization; c) same as b), but with a full amplitude and phase optimization procedure. The FWHM spectral width is about 9.4 nm in a) and c).}
\end{figure}
The first, simple, form of shaping is performed by passing the heralded single-photon pulses through a 10-cm-long BK7 glass block, resulting in their temporal stretching and spectral chirp. We initially impose a simple phase-only polynomial modulation to the spectrum of the LO. Here, it is the (few) coefficients of the polynomial expansion that constitute the genes on which the evolutionary algorithm operates. While the linear component of the spectral phase accommodates for possible temporal delays between the single photon and the LO, the main correction applied by the mask and visible in Fig.3b is in the quadratic phase term, corresponding to the introduction of a linear frequency chirp in the LO pulse. In the next run we let both the spectral phase and amplitude of the shaped LO pulses vary in the adaptive algorithm. In this case a better efficiency is reached, and we obtain the shape of Fig.3c, which still presents the same quadratic spectral phase, but a significantly different spectral intensity that, as expected, is much more similar to that of the un-modulated single-photon.
The best detection efficiency obtained while measuring the dispersed single photons with the optimally-shaped LO pulses reaches $\eta_{opt}\approx 60\%$, whereas using a transform-limited LO pulse reduces it to less than $50 \%$. If, instead of a short glass block, one had propagated the single photons through the dispersive transmission line constituted by a long optical fiber, the resulting quantum state would have been essentially lost to a detection system not taking these S/T modulations into account.

More complex modulations on the profile of the single photon are obtained by shaping the 400 nm pulses pumping the parametric down-conversion crystal. Indeed, it was demonstrated that the single photon conditionally generated in this way essentially inherits the spectral properties of the pump if the filtering of the herald idler mode is sufficiently tight~\cite{aichele02,bellini03,viciani04}. In this case, we place a Michelson interferometer (MI) in the path of the pump pulses to the parametric crystal; the MI splits the pump pulse and recombines the two copies with some adjustable relative delay that causes a sinusoidal modulation of the spectral amplitude with a spectral period inversely proportional to this delay. We use two different configurations of the MI, with approximately the same delay (about 150 fs) and a mod$(2\pi)$ temporal phase shift $\varphi$ of 0 or $\pi$ between the pump pulses. This results in placing the peak or the valley of the sinusoidal modulation in correspondence to the maximum of the pump spectral profile. In one case ($\varphi=0$) the MI essentially acts as a spectral filter that narrows the pump spectrum, whereas in the other ($\varphi=\pi$) it digs a hole in the middle of the pump spectrum, thus producing a two-peaked configuration.
\begin{figure}[h]
\includegraphics [width=1\linewidth]{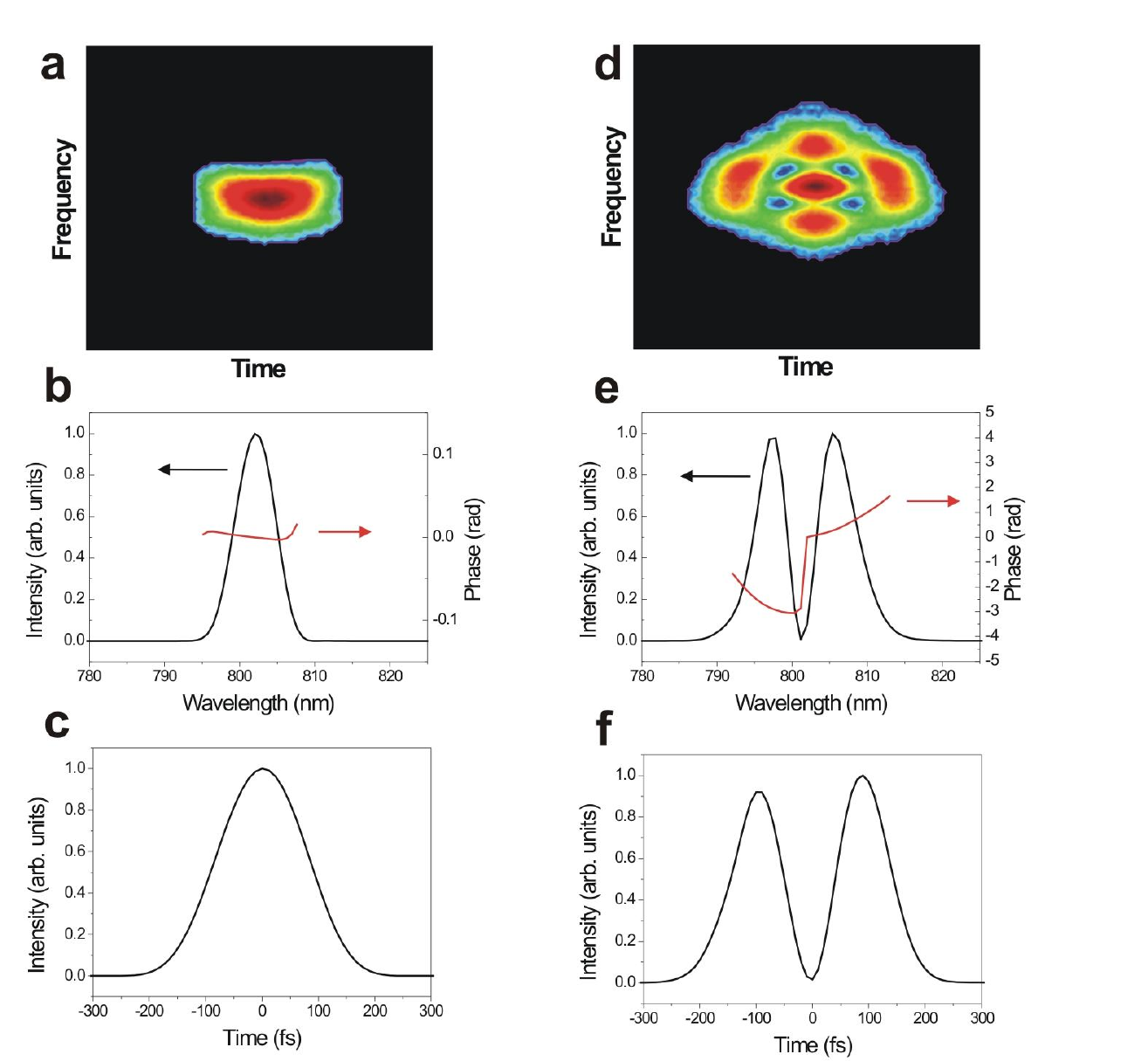}
\caption{\textbf{Effects of a shaped pump pulse}. Measured FROG traces and reconstructed single-photon intensity/phase profiles vs. wavelength and time for (a,b,c) $\varphi=0$ and (d,e,f) $\varphi=\pi$. The measured FWHM pulse duration and spectral width for the $\varphi=0$ case are 178 fs and 6.0 nm, respectively. When $\varphi=\pi$ we observe the expected $\pi$ phase jump in the center of the spectrum. In this case we also find an additional quadratic phase modulation that is likely connected to an observed slight change of the pump spatial mode. Correspondingly, the maximum efficiency achieved with the evolutionary algorithm is limited to about 45$\%$ in this situation.}\label{fig_ph}
\end{figure}
In the $\varphi=0$ situation, the single photon is also shown to acquire the expected narrow spectrum with a relatively flat spectral phase; correspondingly, the single-photon pulse is temporally stretched (see Figs.4a, b, and c). In the $\varphi=\pi$ case, the MI splits the pump into two time-delayed pulses that partially destructively interfere in the overlap region. The double-peak temporal and spectral structure of the pump pulse is then transferred to the heralded single-photon, whose experimentally-reconstructed Wigner function is shown in Fig.5f, and is evident in the reconstructed spectral and temporal profiles of Fig.4e and f. In both cases, starting from average initial values smaller than 10\% for the fitness of the individuals in the first generation of the algorithm, we observe a typically rapid increase of the efficiency over the next few generations followed by a slower adjustment toward the optimum shape (see Fig.2b).

In the latter case, from both the spectral and temporal points of view, the single photon does not occupy just one peak or the other, but rather exists in a coherent superposition over both the two peaks. The coherent delocalization in time and frequency may thus allow one to re-define new spectro-temporal modes $\Psi_1$ and $\Psi_2$ based on these peaks, and therefore use them to encode, manipulate, and detect, qubit information with a single photon. One can, for example, assign the $\ket{1}_{\Psi_1}$ state to the photon occupying the first spectral peak, and the $\ket{1}_{\Psi_2}$ state to the photon in the second one. In this perspective, such a pump modulation allows the realization of the $2^{-1/2}(\ket{1}_{\Psi_1}\ket{0}_{\Psi_2}+\ket{0}_{\Psi_1}\ket{1}_{\Psi_2})$ single-photon qubit. In order to prove that this is indeed the case, instead of just a statistical mixture of the photon occupying one peak or the other, we use the SLM to set the LO in the orthogonal spectral mode obtained by adding a $\phi_{LO}=\pi$ phase shift between the two spectral peaks (see Fig.5b). This corresponds to performing a homodyne measurement of the state by projecting it onto the orthogonal $2^{-1/2}(\ket{1}_{\Psi_1}\ket{0}_{\Psi_2}-\ket{0}_{\Psi_1}\ket{1}_{\Psi_2})$ one. As expected, we find a vanishing efficiency in this case (see the reconstructed Wigner function of the corresponding vacuum state in Fig.5g) and, if the relative phase shift $\phi_{LO}$ is continuously varied, we clearly observe the corresponding cosinusoidal $\eta$ modulation (see Fig.5e), a convincing proof of the coherent nature of the superposition state.
\begin{figure}[h]
\includegraphics [width=1\linewidth]{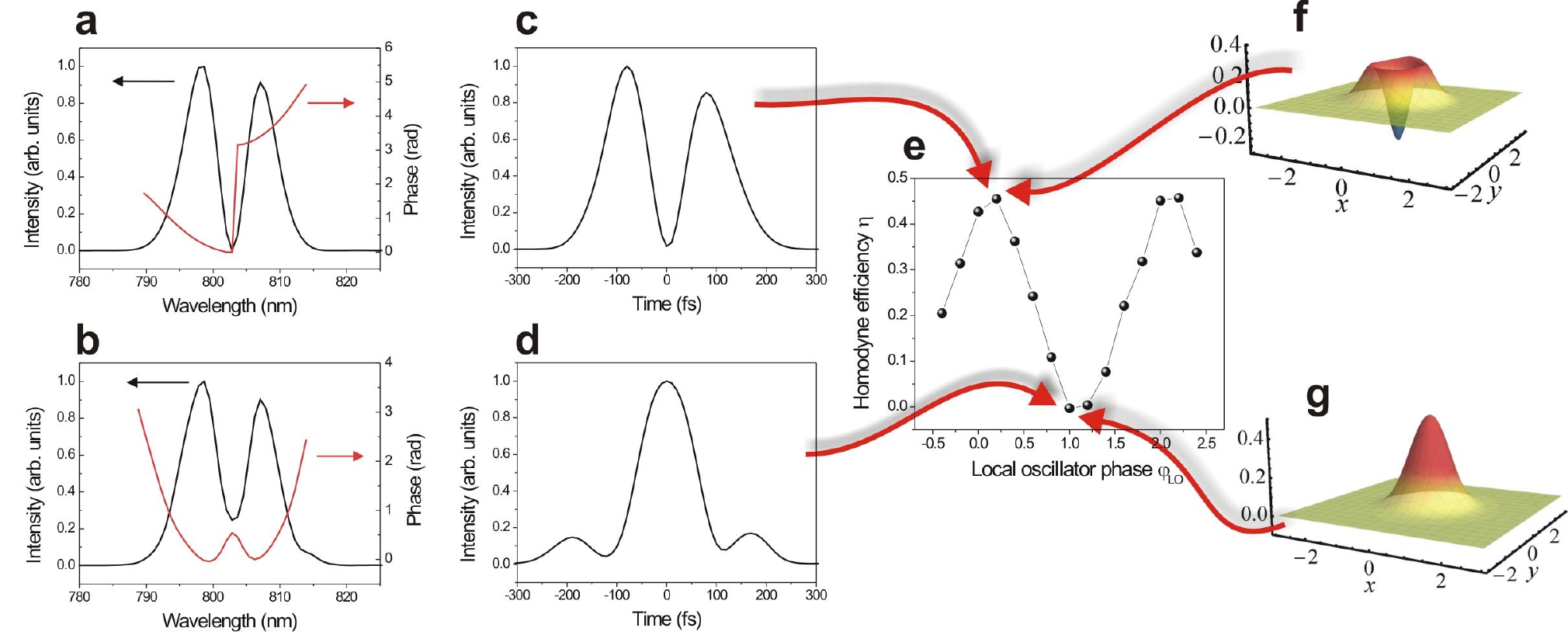}
\caption{\textbf{Probing a spectrally- and temporally-delocalized single photon}. FROG-measured spectral intensity and phase profiles of the LO mode for: a) $\phi_{LO}=0$; b) $\phi_{LO}=\pi$. The corresponding temporal intensity profiles are shown in c) and d), respectively. Note the clear double-peak structure of the LO in the $\phi_{LO}=0$ case and the nearly orthogonal temporal shape obtained in the $\phi_{LO}=\pi$ condition. e) Variation of the measured single-photon homodyne efficiency $\eta$ as a function of the phase $\phi_{LO}$ between the two LO spectral peaks. Reconstructed Wigner functions of the detected state: a single photon in f) for $\phi_{LO}=0$, and vacuum in g) for $\phi_{LO}=\pi$.
}
\end{figure}

The above discussion can be readily extended to higher-dimensional spectral qudits. For example, in the current setup, by simply increasing the MI time delay between the two pump pulses. While the generated single photon is still delocalized between just two temporal modes, in the frequency domain it breaks up coherently in a series of equidistant spectral peaks with a comb-like structure. These distinct spectral modes correspond to the maxima of the sinusoidal modulation of the pump spectrum and their number is roughly limited by the ratio of pump to idler-filter bandwidth. Each spectral mode can be individually analyzed by our homodyne detection scheme with an appropriately-shaped LO, and also their mutual coherence can be completely probed by extending the experimental approach described above. In addition, one might use it to probe the intrinsic multi-Schmidt-mode character of quantum states naturally occurring in parametric down-conversion pumped by short pulses~\cite{law00,wasilewski06,christ11}, which have also been proposed to support multimode quantum information processing~\cite{zhang07,lijie08,dilorenzopires10,neves07}.
Finally, it should be noted that homodyne measurements with a shaped LO might not only be used in the detection stage, but also in the conditional preparation of such multi-mode single-photon states.

In conclusion, we have demonstrated the possibility of combining advanced techniques of ultrafast and quantum optics to measure quantum states of light in unknown S/T modes, in a way that may allow one to recover ``unreadable'' quantum information or pre-compensate the possible deformations encountered during propagation in arbitrary dispersive channels. Moreover, the possibility of generating and detecting ultrashort quantum states of light in multiple, arbitrary, S/T modes opens new exciting opportunities by providing access to a much larger Hilbert space for encoding, manipulating, and decoding, quantum information.

\section{Acknowledgments}
A.Z. and M.B. acknowledge support by Ente Cassa di Risparmio di Firenze, Regione Toscana under project CTOTUS, and EU under ERA-NET CHIST-ERA project QSCALE. K.N.C. acknowledges support by Alexander von Humboldt Foundation and LaserLab Europe.

Correspondence and requests for materials should be addressed to M.B.~(email: bellini@ino.it).

\bibliography{Fock_bib}

\end{document}